\documentclass[prl,superscriptaddress,showpacs,longbibliography,reprint]{revtex4-1}
\usepackage[utf8]{inputenc}
\usepackage{amsmath}
\usepackage{amsfonts}
\usepackage{amssymb}
\usepackage{braket}
\usepackage{dsfont}

\newcommand{\N}{\mathcal{N}}

\newcommand{\id}{\mathds{1}}

\newcommand{\ketbra}[2]{\ket{#1}\hspace{-2.1pt}\bra{#2}}

\DeclareMathOperator{\Tr}{Tr}

\newcommand{\spl}{\sigma^{+}}
\newcommand{\sm}{\sigma^{-}}
\newcommand{\nb}{\overline{n}}

\newcommand{\sx}{\sigma^{x}}
\newcommand{\sy}{\sigma^{y}}

\usepackage{graphicx}

\usepackage{color,soul}

\begin{document}
\title{Asynchronism and nonequilibrium phase transitions\\ in $(1+1)$D quantum cellular automata}

\author{Edward Gillman}
\affiliation{School of Physics and Astronomy, University of Nottingham, Nottingham, NG7 2RD, UK}
\affiliation{Centre for the Mathematics and Theoretical Physics of Quantum Non-Equilibrium Systems,
University of Nottingham, Nottingham, NG7 2RD, UK}

\author{Federico Carollo}
\affiliation{Institut f\"{u}r Theoretische Physik, Universit\"{a}t T\"{u}bingen, Auf der Morgenstelle 14, 72076 T\"{u}bingen, Germany}

\author{Igor Lesanovsky}
\affiliation{School of Physics and Astronomy, University of Nottingham, Nottingham, NG7 2RD, UK}
\affiliation{Centre for the Mathematics and Theoretical Physics of Quantum Non-Equilibrium Systems,
University of Nottingham, Nottingham, NG7 2RD, UK}
\affiliation{Institut f\"{u}r Theoretische Physik, Universit\"{a}t T\"{u}bingen, Auf der Morgenstelle 14, 72076 T\"{u}bingen, Germany}

\begin{abstract}
Probabilistic cellular automata provide a simple framework for the exploration of classical nonequilibrium processes. Recently, quantum cellular automata have been proposed that rely on the propagation of a one-dimensional quantum state along a fictitious discrete time dimension via the sequential application of quantum gates. The resulting $(1+1)$-dimensional space-time structure makes these automata special cases of feed-forward quantum neural networks. Here we show how asynchronism --- introduced via non-commuting gates --- impacts on the collective nonequilibrium behavior of quantum cellular automata. We illustrate this through a simple model, whose synchronous version implements a contact process and features a nonequilibrium phase transition in the directed percolation universality class. Non-commuting quantum gates lead to an ``asynchronism transition", i.e.~a sudden qualitative change in the phase transition behavior once a certain degree of asynchronicity is surpassed. Our results show how quantum effects may lead to abrupt changes of non-equilibrium dynamics, which may be relevant for understanding the role of quantum correlations in neural networks.
\end{abstract}

\maketitle
\begin{figure}[t]
\centering
\includegraphics[width=1\linewidth]{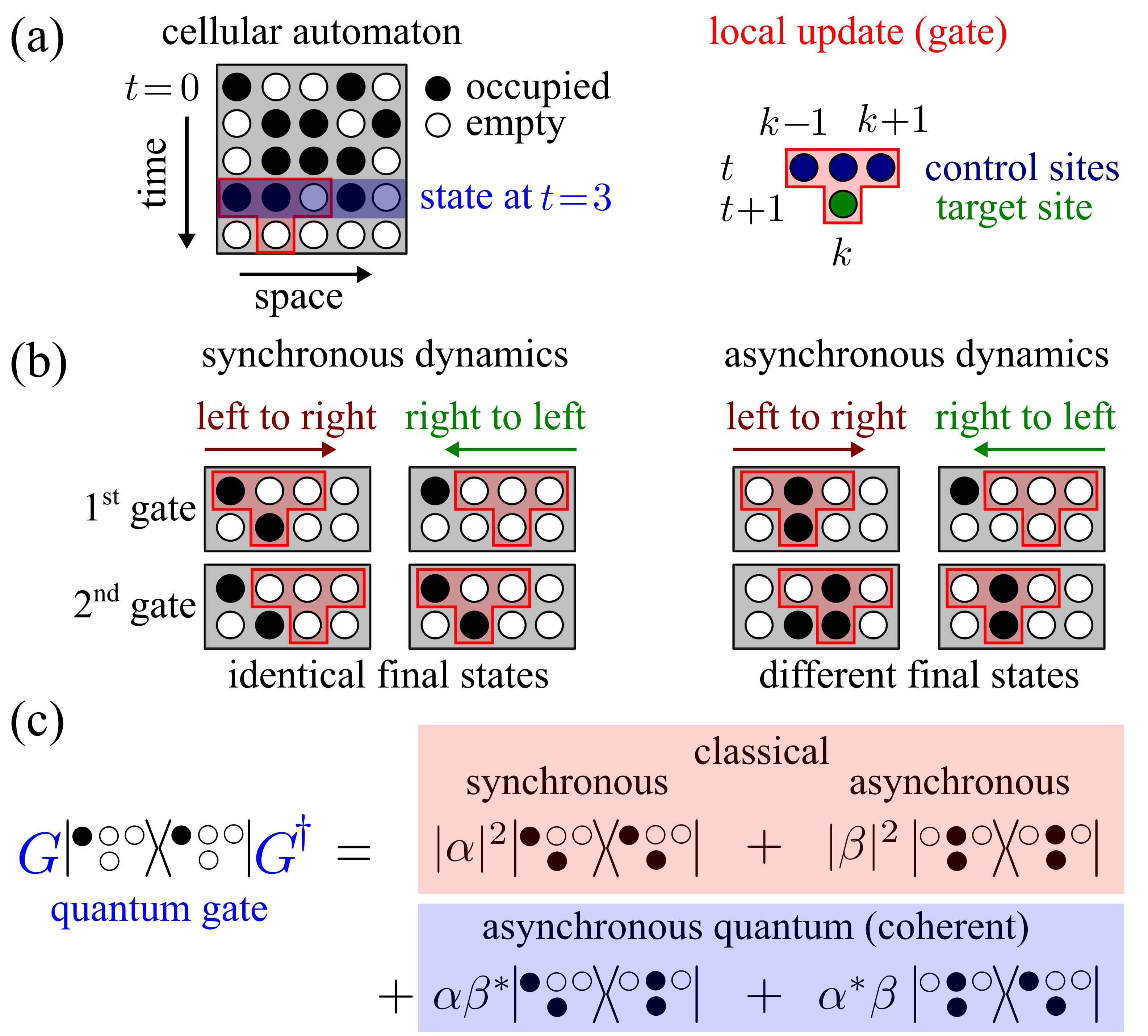}
\caption{\textbf{Asynchronism in classical CA and $(1+1)$D QCA:} (a) A CA consists of a $2$D lattice of two-level systems, which can either be in an occupied or empty state. The vertical dimension of the lattice provides an effective discrete-time dimension and propagation along the time direction is achieved through the sequential application of local gates. These perform operations on a target site at row $t+1$ and in position $k$ according to the state of three control sites (at position $k-1,k,k+1$) in the previous row. After the row $t+1$ has been completely updated, its state represents the state of an effective $1$D system at time $t+1$. \textbf{(b)} During a single time update, if the local gates do not modify the control sites, the order of their application is irrelevant (synchronous dynamics). If local gates change the control sites, the order of their application is relevant and can produce different final states (asynchronous dynamics). \textbf{(c)} While in a classical CA, the occupation of a target site solely depends on occupation probabilities  of control sites, asynchronism in a QCA is linked to quantum coherent processes. In the panel, it is shown how an asynchronous gate can generate, in addition to classical asynchronous terms, a coupling between the occupation in the target site and coherence in the control sites. 
}
\label{Fig1}
\end{figure}

Nonequilibrium processes can display collective effects and critical behavior. In the vicinity of nonequilibrium phase transitions (NEPTs), the resulting phenomenology can show macroscopic features that are shared by different mathematical and physical models. This so-called universality allows for diverse systems to be gathered into few universality classes, and enables the investigation of general properties of emergent phenomena through the analysis of minimal models within a class \cite{Henkel2008,Henkel2010}. In the study of classical systems, a paradigmatic setting for exploring nonequilibrium universality is that of  $(1+1)$D cellular automata (CA). These consist of $2$D lattice models realising an effective $1$D system discrete-time dynamics, as shown in Fig.~\ref{Fig1}(a). The propagation of the $1$D state from time $t$ to time $t+1$ takes place through the application of a sequence of local gates (or rules) that operate on the (target) row $t+1$, controlled by the state of row $t$, see Fig.~\ref{Fig1}(a). %The row $t+1$ of the lattice then provides the state of the effective $1$D system at time $t+1$. 
Such a dynamics can either be deterministic, usually implemented through unitary gates, or probabilistic, with non-unitary local updates. In the latter case, by suitably choosing the gates, these automata provide discrete-time versions of continuous-time dynamics. Owing to their simple structure, this has allowed for a deep understanding of several nonequilibrium processes \cite{Henkel2008,Henkel2010,Domany1984, Bagnoli2001,Bagnoli2014,Hinrichsen2000,Lubeck2005}.

Recently, quantum versions of these automata have been introduced and dubbed $(1+1)$D quantum cellular automata (QCA) \cite{Lesanovsky2019,Gillman2020,Gillman2021}. These models naturally include their classical counterparts as a limiting case. For instance, probabilistic cellular automata (PCA) can be reproduced through unitary quantum gates, leading to a quantum state whose diagonal elements provide the configuration probabilities of the associated classical CA. However, in these settings the system's state also displays non-classical properties, such as superposition and coherence \cite{Lesanovsky2019,Gillman2021b}. The study of $(1+1)$D QCA is appealing for at least two reasons. Firstly, QCAs can be realized on current quantum simulation platforms, such as two-dimensional Rydberg lattice gases \cite{zeiher2016,kim2018,browaeys2020,ebadi2020}. Secondly, while closely linked to unitary $1$D QCA, as studied, e.g., in Refs. \cite{Wiesner2009,Cirac2017,Arrighi2019,Farrelly2020,Hillberry2021}, $(1+1)$D QCA are in fact equivalent to feed-forward quantum neural networks applied in quantum machine learning \cite{Beer2020}. $(1+1)$D QCA thus allow for the analysis of how quantum dynamical processes alter nonequilibrium dynamics in these structures, and more generally enable the exploration of the impact of non-classical effects on emergent universal behavior. 

In this paper, we show that a key property for defining the ``quantumness" of a QCA is {\it asynchronism}, which occurs when the automaton's dynamics depends on the order of the application of the gates [see sketch in Fig.~\ref{Fig1}(b)]. In fact, asynchronism itself is not necessarily a quantum feature, and has also been investigated in classical settings \cite{Boure2012,Bandini2012,Fates2013}. However, in QCA, asynchronous --- i.e. non-commuting --- gates can generate a dependence between diagonal observables of one time slice and coherence observables in the previous one, see Fig.~\ref{Fig1}(c). This coupling between diagonal populations and off-diagonal coherences can ultimately lead to a non-analytic change in the critical behavior of the system. We illustrate this effect by considering a paradigmatic nonequilibrium system, the so-called {\it contact process} (CP) \cite{Harris1974,Henkel2008}, which we encode in a synchronous QCA. %Modifying the elementary gate to allow for a controllable degree of asynchronism. We show that, 
While for low asynchronism the mean-field analysis of $(1+1)$D QCA displays a second-order NEPT in the directed percolation (DP) universality class --- as expected from the corresponding PCA \cite{Domany1984} --- there exists a critical value of asynchronism above which the NEPT of the model qualitatively changes and becomes first-order. We thus term this an ``asynchronism transition". 

A similar change of collective behavior upon the introduction of quantum effects was also recently reported in continuous-time quantum versions of classical nonequilibrium models, studied in Refs. \cite{Marcuzzi2016,Buchhold2017,Roscher2018,Carollo2019,Gillman2019,Jo2021}, and we show that this phenomenon is indeed related with the ``asynchronism transition" in our $(1+1)$D QCA.  %Similarly to asynchronism in our QCA, the change in the universal physics of these models occurs when the ratio of quantum effects relative to classical processes surpasses a certain threshold. We explain this similarity by showing that asynchronism in our QCA generates dynamical terms resembling those of the quantum contact process (QCP) Hamiltonian, and that the strength of asynchronism is related to the relative strength of quantum effects. 
Our results shed light on the role of quantum effects in nonequilibrium settings and their impact on emergent collective behavior. Moreover, given the connection between asynchronism and universal computation in feed-forward quantum neural networks \cite{Beer2020}, our work also offers an interesting link between quantum many-body dynamics and quantum machine learning.

\noindent \textbf{Synchronous $\mathbf{(1+1)}$D QCA.} In the $2$D lattice of the QCA each site can be in an empty $\ket{\circ}$ or occupied $\ket{\bullet}$ state, see Fig. \ref{Fig1}(a). The initial state of the lattice, $\ket{\psi_0}$, is chosen as a product state, where the row $t=0$ contains the initial $1$D configuration, while the sites in the remaining rows are initialized in state $\ket{\circ}$. The full $2$D lattice dynamics proceeds iteratively as $\ket{\psi_{t+1}}=\mathcal{G}_{t}\ket{\psi_{t}}$, where $\mathcal{G}_t$ acts on rows $t$ and $t+1$. 
This global update is composed of an ordered product of local gates, $G_{t,k}$, which act to update a target site at $(t+1,k)$, with one gate per target site. For example, choosing a left-to-right ordering one has $\mathcal{G}_{t} = \dots G_{t,k}\dots G_{t,2}G_{t,1}$. The time-evolved $1$D system state is obtained as $\rho_t=\mathrm{Tr}'\left(\ket{\psi_{t}}\!\bra{\psi_{t}}\right)$, with the trace taken over all sites except for those in row $t$ [cf.~Fig.\ref{Fig1}(a)].

The simplest local unitary gate is of the form
\begin{align}
G_{t,k} = \sum_{\N} P_{\N} \otimes U_{\N} ~.
\label{eqn:basic_gate_form}
\end{align}
Here, $\N$ labels the basis states of the ``control" sites in the neighbourhood of site $k$ on row $t$. For example, a three-site neighbourhood has $8$ basis states, $\N=({\circ\!\circ\!\circ}, \circ\!\circ\!\bullet, ..., \bullet\!\bullet\!\bullet$). The unitary operator $U_{\N}$ ``rotates'' the target (on row $t+1$ and in position $k$) conditioned on the state of the control sites. This is enforced by the action of the projector $P_{\N} = \ketbra{\N}{\N}$ onto these sites. Here and throughout, we will use the symbol $\otimes$ to separate control sites (to the left) and target sites (to the right). Since in Eq.~\eqref{eqn:basic_gate_form} only orthogonal projectors act on control sites, gates $G_{t,k}$ acting on different target sites commute. The order of their application is thus irrelevant and the dynamics is synchronous, i.e., all sites can be updated simultaneously [see Fig. \ref{Fig1}(b)]. 

To illustrate that such synchronous $(1+1)$D QCA allows for the implementation of a range of canonical nonequilibrium models \cite{Wolfram1983,Wolfram2002,Bagnoli2001,Bagnoli2014} we consider the realization of the so-called contact process \cite{Domany1984,Hinrichsen2000}. The contact process features three elementary ingredients: decay, i.e. the transition of a site from occupied to empty ($\bullet\rightsquigarrow\circ$); coagulation, which is also the transition of a site from full to empty but facilitated (conditioned) by one of its neighbors ($\bullet\bullet\rightsquigarrow\bullet\circ$); and branching, which is facilitated excitation of the form $\bullet\circ\rightsquigarrow\bullet\bullet$. Note that the contact process possesses the absorbing state $...\circ\!\circ\!\circ...$ from which the above-mentioned dynamical processes allow no escape. Whether this absorbing state is reached at stationarity depends on the values of the rates (or probabilities), with which the elementary processes occur. An instance of the contact process on a $(1+1)$D QCA is realized by the gate
\begin{eqnarray}
 G_{t,k} & = &\Pi_{k}n_k \otimes U_{\circ\bullet\circ} +\Pi_{k}\nb_k \otimes \id \nonumber  \\
 &&+ \bar{\Pi}_{k} n_k \otimes U_{\bullet} + \bar{\Pi}_{k}\nb_{k} \otimes U_{\circ},
\label{eqn:classical_process_gate}
\end{eqnarray}
which has the form of Eq.~\eqref{eqn:basic_gate_form}. Here $n_k = \ketbra{\bullet}{\bullet}_k$ and $\nb_k = \ketbra{\circ}{\circ}_k=\id - n_k$ project onto the occupied and empty state of site $k$, respectively. Furthermore, we have defined the projectors $\Pi_{k}=\bar{n}_{k-1}\bar{n}_{k+1}$ and their complements $\bar{\Pi}_{k} = \id - \Pi_{k}$. The unitaries $U_{\alpha}$, with labels $\alpha=(\circ\!\bullet\!\circ,\bullet,\circ)$, perform a (coherent) flip of the target site, which is conditioned on the state of the controls. Note that the first unitary considers the case of empty left/right control sites, while the latter two unitaries  act on the target only if at least one of the left/right control sites is occupied. They are parametrized as $U_{\circ\bullet\circ} = \sqrt{p_{\circ\bullet\circ}}\, \id - i \sqrt{q_{\circ\bullet\circ}}\, \sx$ and $U_{\circ/\bullet}   = \sqrt{q_{\circ/\bullet}}\, \id - i \sqrt{p_{\circ/\bullet}}\, \sx$ with $\sx = \ketbra{\bullet}{\circ} + \ketbra{\circ}{\bullet}$. The parameters $q_{\circ\bullet\circ}$ and $p_{\circ/\bullet} \in [0,1]$ are the flipping probabilities, and $q_\alpha = 1 - p_\alpha$.

In the gate in Eq.~\eqref{eqn:classical_process_gate}, the control sites have been separated by singling out the central one, so that we can associate to the target site [in position $(t+1,k)$] a specific control site [the one in position $(t,k)$], which we regard as its  ``past". This allows for the mean occupation number, $\langle n_k\rangle_{t+1}$, of the target to be calculated iteratively as,
\begin{eqnarray}
\langle n_k\rangle_{t+1} &=&  q_{\circ\bullet\circ} \langle \Pi_{k} n_{k}\rangle_t + p_{\bullet} \langle \bar{\Pi}_{k} n_k\rangle_t + p_{\circ} \langle \bar{\Pi}_{k} \bar{n}_{k}\rangle_t \label{eqn:full_qca_ccp_mfe}\\
&\approx& q_{\circ\bullet\circ} \langle \Pi_{k} \rangle_t \langle n_{k}\rangle_t + p_{\bullet} \langle \bar{\Pi}_{k} \rangle_t \langle n_k\rangle_t + p_{\circ} \langle \bar{\Pi}_{k} \rangle_t \langle\bar{n}_{k}\rangle_t\nonumber,
\end{eqnarray}
where we performed a mean-field decoupling in the second line \cite{SM}. This form makes the interpretation of the probabilities entering the unitaries $U_\alpha$ rather transparent: $q_{\circ\bullet\circ}$ is the probability that the target site $k$ gets occupied given that the control site $k$ is occupied while its neighbors are empty. Since the occupation number can only decrease under this process this effectively implements $\bullet\rightsquigarrow\circ$. The probability $p_{\bullet}$ is the probability of having an occupied target when there is at least one of the external controls and the central one occupied. This also describes a decay process, but here in combination with the so-called {\it coagulation process}, i.e.~the annihilation of two adjacent occupied sites, e.g. $\bullet\bullet\rightsquigarrow\bullet\circ$. Finally, the probability $p_{\circ}$ parametrizes the strength of a {\it branching process} ($\bullet\circ\rightsquigarrow\bullet\bullet$), i.e.~the creation of an excitation in an empty site facilitated by the presence of a neighboring occupied site. All these elementary ingredients combined yield the contact process \cite{Hinrichsen2000}. Finally, by taking the continuous-time limit of Eq.~\eqref{eqn:full_qca_ccp_mfe}, which entails the expansion $\langle n_k\rangle_{t+1} \approx \langle n_k\rangle_{t}+\Delta t \frac{d}{dt}\langle n_k\rangle_{t}$, with small time step $\Delta t$, one obtains a continuous-time contact process \cite{Hinrichsen2000} with coagulation rate $\kappa_{\mathrm{c}}=(q_{\circ\bullet\circ}-p_\bullet)/\Delta t$, branching rate $\kappa_{\mathrm{b}}=p_{\circ}/\Delta t$ and decay rate $\gamma=p_{\circ\bullet\circ}/\Delta t$ \cite{SM}. 

\noindent \textbf{Asynchronous $\mathbf{(1+1)D}$ QCA.} The discrete-time dynamics in Eq.~\eqref{eqn:full_qca_ccp_mfe} is classical in the sense that it only connects diagonal observables. A natural question is: what is the minimal modification that we can make to the gate $G_{t,k}$ in a way that diagonal observables at time $t+1$ depend on coherence at the previous time slice? As sketched in Fig.~\ref{Fig1}(b-c), we achieve this through asynchronism introduced by non-commuting gates.

To break the commutativity of adjacent gates, we consider terms that can modify the state of the control sites along with the target one, see Fig. \ref{Fig1}(b). In general, this is accomplished by gates of the form 
\begin{align}
G_{t,k} = \sum_{\N} P_{\N} \otimes U_{\N} + \sum_{\N \neq \N'} \ketbra{\N}{\N'} \otimes O_{\N, \N'} ~,
\label{eqn:full_gate_form}
\end{align}
where the unitarity of $G_{t,k} $ imposes constraints on the operators $O_{\N, \N'}$. Since our aim is to generate a dependence of $\langle n_k\rangle_{t+1}$ on coherence observables for the control site $k$, we focus on transition operators $\ketbra{\N}{\N'}$ that solely modify this control, i.e., $\ketbra{\N}{\N'} = \ketbra{\N}{\N}\sigma^{\pm}_k$, where $\spl = \ketbra{\bullet}{\circ}$ and $\sm = \ketbra{\circ}{\bullet}$. For concreteness, we choose here the gate
\begin{eqnarray}
 G_{t,k} &= &\Pi_k n_{k} \otimes U_{\circ\bullet\circ} + \Pi_k \nb_{k} \otimes \id \nonumber  \\
 &&+ \sqrt{1-\lambda}\, \left[ \bar{\Pi}_k n_{k} \otimes U_{\bullet} + \bar{\Pi}_k \nb_{k} \otimes U_{\circ} \right] \nonumber \\
 &&+ \sqrt{\lambda}\, \bar{\Pi}_k \left[ \spl_{k} \otimes U_{\bullet} U_{+} -  \sm_k \otimes U_{\circ} U_{+}^{\dagger}\right] ~,
\label{eqn:quantum_process_gate}
\end{eqnarray}
which depends on the further unitary $U_{+} = i \sqrt{q}\, \id - \sqrt{p}\, \sx $. In Eq.~\eqref{eqn:quantum_process_gate}, the parameter $\lambda \in [0,1]$ controls the strength of the asynchronism. When $\lambda = 0$ we have the synchronous case considered before in Eq.~\eqref{eqn:classical_process_gate}. As $\lambda$ is increased the gates acting on adjacent target sites do not commute, with the size of the commutator -- quantified by the matrix norm -- increasing with $\lambda$. Considering the analogue of Eq.~\eqref{eqn:full_qca_ccp_mfe} for the gate in Eq.~\eqref{eqn:quantum_process_gate}, we find after a mean-field decoupling \cite{SM},
\begin{eqnarray}
\langle n\rangle_{t+1} &=  &r_{\circ\bullet\circ} \langle \Pi_k\rangle_t \langle n_{k}\rangle_t + r_{\bullet} \langle \bar{\Pi}_k\rangle_t \langle n_{k}\rangle_t + r_{\circ} \langle \bar{\Pi}_{k}\rangle_t \langle \bar{n}_{k}\rangle_t \nonumber \\
&&+r_{\ast} \langle \bar{\Pi}_{k}\rangle_t \braket{\sigma^{y}_k}_{t}, 
\label{eqn:full_qca_mfe}
\end{eqnarray}
with $\sy = -i\ketbra{\bullet}{\circ} +i \ketbra{\circ}{\bullet}$. The coefficients are 
\begin{align}    
    r_{\circ\bullet\circ}&= q_{\circ\bullet\circ} , \nonumber \\
    r_{\bullet}&=(1-\lambda)p_\bullet+\lambda \left(\sqrt{p_\circ}\sqrt{q}+\sqrt{p}\sqrt{q_{\circ}}\right)^{2} , \nonumber \\
    r_{\circ}&=  (1-\lambda) p_{\circ}+\lambda (\sqrt{p_\bullet}\sqrt{q}-\sqrt{p}\sqrt{q_\bullet})^{2} , \nonumber \\
   r_{\ast}&=  \sqrt{\lambda}\sqrt{1-\lambda}[\sqrt{q}\left(p_\bullet+p_{\circ}\right) \nonumber \\
    &~~~~~~~~~~~~~~+\sqrt{p}\left(\sqrt{p_\circ}\sqrt{q_{\circ}}-\sqrt{p_\bullet}\sqrt{q_\bullet}\right)].
\label{eqn:QCP_QCA_processes}
\end{align}
Crucially, we see that this equation connects the density operator $n$ of the target site with the coherence observable $\sigma^y$ for the central control, which, as mentioned before, we interpret as the ``past'' of the target. Only when $r_{\ast} = 0$ is the equation closed on diagonal observables. 

\begin{figure}[t]
\centering
\includegraphics[width=1\linewidth]{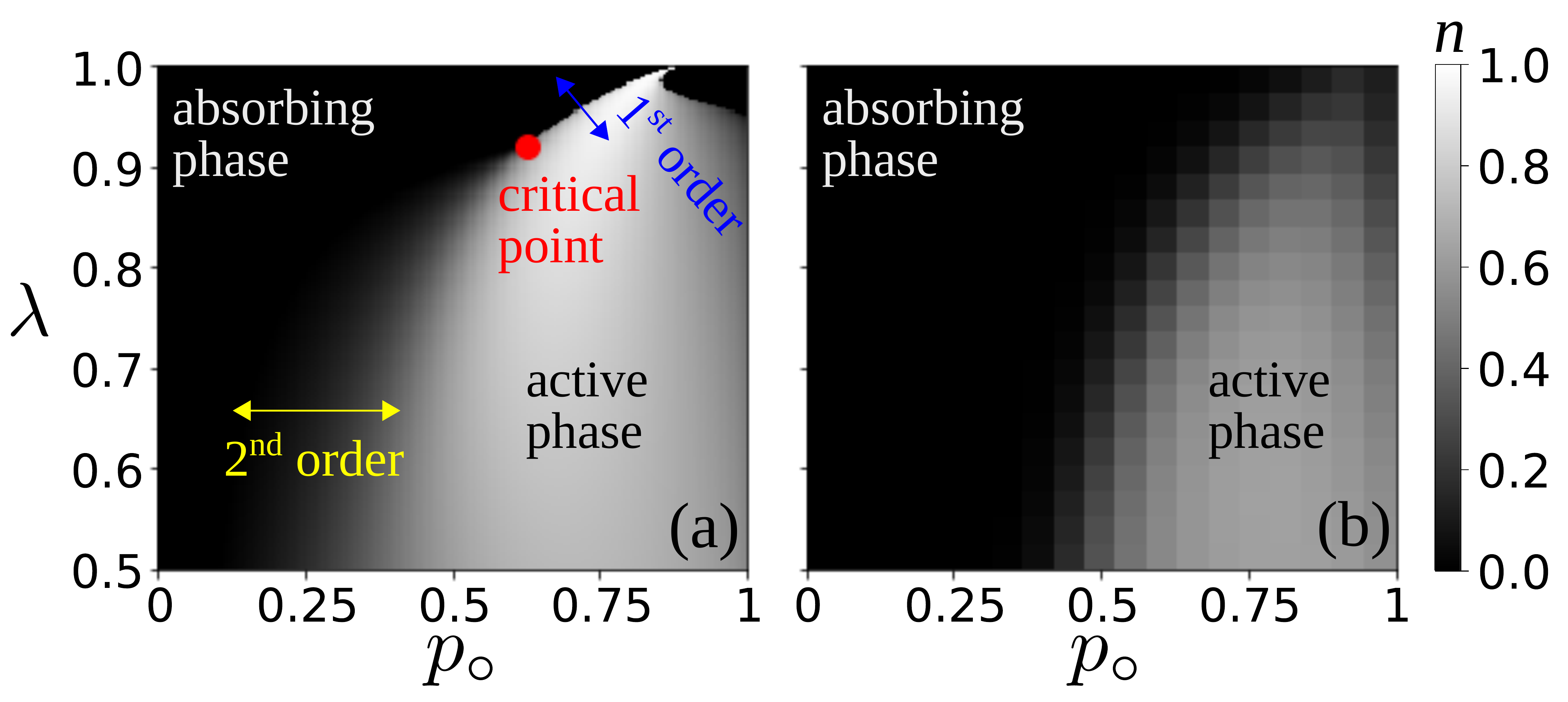}
\caption{\textbf{Nonequilbrium phase transition in $(1+1)$D QCA:} (a) Stationary phase diagram for the model described by the gate \eqref{eqn:quantum_process_gate}. An absorbing state phase transition is displayed as a function of the asynchronism parameter $\lambda$ and the branching probability $p_{\circ}$. The stationary density is estimated by performing $1000$ iterations of the mean-field equations \cite{SM}. For strong asychronicity, i.e. $\lambda \gtrsim 0.92$, the phase transition changes from continuous (in the directed percolation universality) to discontinuous.  (b) Phase diagram obtained for a $(1+1)$D QCA using tensor networks \cite{Gillman2021b}. Here, the density is calculated using bond-dimension $\chi = 64$, lattice size $L=64$, and by iterating over $100$ time steps.}
\label{fig:phase_diagrams}
\end{figure}

\noindent \textbf{Asynchronism transition.} To assess the impact of asynchronism on our $(1+1)$D QCA, we investigate the stationary state within a mean-field assumption \cite{SM}. For the following analysis we fix the parameters $q_{\circ\bullet\circ} = p_{\bullet} = p = 0.1$. As can be seen in Fig. \ref{fig:phase_diagrams}(a), for any given value of $\lambda$ the QCA displays an NEPT from the absorbing state with all empty sites to a state possessing a finite density of occupied sites, i.e. $\langle n\rangle_\infty \neq0$. The critical curve separating those two phases can be parametrized by the strength of asynchronism, $\lambda = \lambda_{\mathrm{c}}(p_{\circ})$. For $\lambda=0$ (not shown in the panel) the QCA coincides with a discrete-time version of the contact process. As such, it shares with it a continuous phase transition in the DP universality class. This continuous transition persists when increasing $\lambda$. However, for large enough $\lambda$, we observe a dramatic change: at $\lambda^{*} \approx 0.92$ the phase transition becomes first-order and thus no longer falls into the DP universality class. 

Since this  change in the universal physics occurs for increasing $\lambda$ along the critical curve $\lambda_{\mathrm{c}}$, we will refer to this as an ``asynchronism transition". 
Going beyond the mean-field level, tensor network methods can be used to estimate the phase diagram \cite{Gillman2021b}, see Fig.~\ref{fig:phase_diagrams}(b). Qualitative agreement with the mean-field is found, although the NEPT appears to be continuous throughout. In fact, this behaviour is similar to that of the so-called quantum contact process  \cite{Carollo2019,Gillman2019,Jo2021}, as we discuss in the following.

\noindent \textbf{Relation to the quantum contact process.} 
The quantum contact process (QCP) is a continuous-time Markovian open quantum system which features the same processes as the contact process, and also includes an additional coherent term, so-called quantum branching ($\bullet\circ \leftrightarrow \bullet\bullet$) with rate $\Omega$ \cite{Marcuzzi2016,Buchhold2017}. This process is implemented by a quantum Hamiltonian, which describes constrained (Rabi) oscillations at site $k$ that can only occur if the neighbours of $k$ are not in the empty state. Just as the classical contact process, the QCP displays a phase transition from an absorbing state to an active phase. The universality class is, however, in general different. At the mean-field level  this change of universal behavior becomes manifest by the fact that beyond a certain critical ratio $g^{*}$ of quantum and classical branching rates, $g = \Omega/\kappa_{\mathrm{b}}$, the transition changes from being continuous to first-order \cite{Marcuzzi2016,Buchhold2017,Roscher2018}. Beyond mean-field theory, numerical simulations show that, in $1$D, the exact phase transition does not change from continuous to first-order. However, there is still a critical value of $g$ below which the universality class of the model is DP, and above which it is not \cite{Carollo2019,Gillman2019,Jo2021}. Thus, the QCP displays a transition in its universal physics analogous to that of the QCA considered previously, with the mean-field transition also signalling a corresponding transition in the exact model. 

\begin{figure}[t]
\centering
\includegraphics[width=1\linewidth]{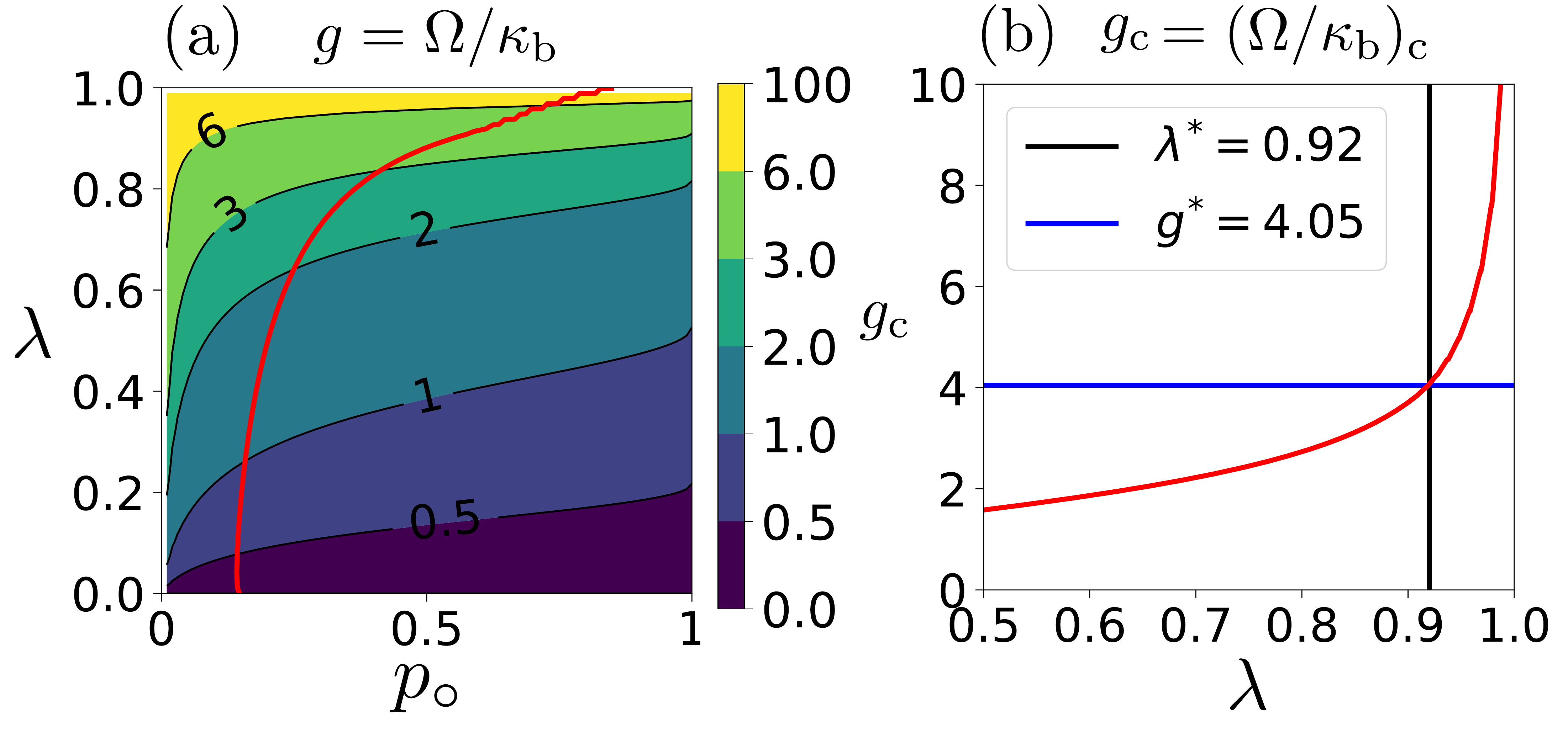}
\caption{\textbf{Quantum and classical processes in asynchronous $(1+1)$D QCA:} (a) By equating Eqs. \eqref{eqn:QCP_QCA_processes} and \eqref{eqn:qcp_processes}, the relative strength of quantum to classical branching, $g = \Omega/\kappa_{b}$, can be examined for the gate in Eq.~\eqref{eqn:quantum_process_gate}. As can be seen, $g$ (shown as lines of constant value, with coloured areas indicating regions between these values) increases monotonically with the strength of asynchronism in the QCA, parametrised by $\lambda$. The solid red line indicates the critical curve of the QCA, estimated by taking the line of constant $n = 0.1$ in the mean-field phase diagram of Fig.~\ref{fig:phase_diagrams}(a). (b) The behaviour of $g$ along the critical line, denoted as $g_{c}$ here, is shown for $\lambda \in [0.5, 1]$, and displays a rapid increase with $\lambda$. The critical point of the asynchronism transition, $\lambda^{*}$, can be identified with a critical value of $g^{*} = 4.05$.}
\label{fig:QCP_params}
\end{figure}

The close resemblance between the phenomenology of our QCA and that of the QCP, suggests that the inclusion of asynchronism in the QCA in fact introduces a microscopic process akin to quantum branching \cite{Marcuzzi2016,Buchhold2017,Roscher2018,Carollo2019,Gillman2019,Gillman2021b,Jo2019,Jo2020,Jo2021,Nigmatullin2021}. It also suggests that the observed asynchronism transition may be understood in terms of the relative strength of quantum to classical branching. To verify this, we consider the Heisenberg equation for $n$ of the QCP \cite{Marcuzzi2016,Buchhold2017,SM}. Upon discretization with time-step $\Delta t$, this is indeed equivalent to Eq.~\eqref{eqn:full_qca_mfe} with coefficients,
\begin{align}
    r_{\circ\bullet\circ} &= 1 - \gamma \Delta t  ~, && r_{\bullet} = 1 - \gamma \Delta t - \kappa_{\text{c}} \Delta t   ~, \nonumber\\
r_{\circ}&=  \ \kappa_{\text{b}} \Delta t~, && r_{\ast} =  \Omega \Delta t~ .
\label{eqn:qcp_processes}
\end{align}

Comparing Eq.~\eqref{eqn:qcp_processes} with Eq.~\eqref{eqn:QCP_QCA_processes}, we  note that removing asynchronism  ($\lambda \to 0$) is equivalent to removing coherent branching ($\Omega\Delta t \to 0$) in the QCP. %This shows that asynchronism in the QCA indeed corresponds to coherent branching in the QCP. 
Next, we re-characterise the asynchronism transition in terms of the processes of the QCP. This is achieved by equating the Eqs.~\eqref{eqn:qcp_processes} and \eqref{eqn:QCP_QCA_processes}, which allows for the definition of the parameter $g = \Omega/\kappa_{\mathrm{b}}$ for our QCA, see Fig. \ref{fig:QCP_params}(a). Clearly, increasing the value of $\lambda$ corresponds to increasing $g$. The critical curve, $\lambda = \lambda_{c}(p_{\circ})$, can then also be parametrised by values of $g$, see Fig. \ref{fig:QCP_params}(b). Evidently, the asynchronism transition in the QCA can be understood in the same manner as the transition in the QCP. In terms of $g$, the critical point of the asynchronism transition, $\lambda^{*}$, is the point $g^{*}$, where quantum branching is sufficiently stronger than classical branching, leading to a change of universal physics.

\noindent \textbf{Conclusion.} We have shown how asynchronism impacts on the nonequilibrium phase transition behavior of $(1+1)$D QCA. Building on the connection between classical probabilistic CA and continuous-time nonequilibrium dynamics \cite{Domany1984,Hinrichsen2000}, we have demonstrated how asynchronism gradually introduces genuine quantum effects to an otherwise classical contact process dynamics, and how this results in a qualitative change of universal behaviour. Due to the connection between QCA and quantum neural networks, this analysis might find application in quantum machine learning, for instance, in providing physical insights into the dynamics of information retrieval and the impact of quantum effects --- e.g.~caused by asynchronism --- on their capability of performing computational tasks.

\textbf{Acknowledgments.} We acknowledge support from EPSRC [Grant No. EP/R04421X/1], from the ``Wissenschaftler R\"{u}ckkehrprogramm GSO/CZS" of the Carl-Zeiss-Stiftung and the German Scholars Organization e.V., as well as through The Leverhulme Trust [Grant No. RPG-2018-181], and the Deutsche Forschungsgemeinschaft through SPP 1929 (GiRyd), Grant No. 428276754, as well as through Grant No. 435696605. We are grateful for access to the University of Nottingham's Augusta HPC service. 

\bibliographystyle{apsrev4-1}
\bibliography{QCA_bib_ab}

%merlin.mbs apsrev4-1.bst 2010-07-25 4.21a (PWD, AO, DPC) hacked
%Control: key (0)
%Control: author (72) initials jnrlst
%Control: editor formatted (1) identically to author
%Control: production of article title (-1) disabled
%Control: page (0) single
%Control: year (1) truncated
%Control: production of eprint (0) enabled
\begin{thebibliography}{37}%
\makeatletter
\providecommand \@ifxundefined [1]{%
 \@ifx{#1\undefined}
}%
\providecommand \@ifnum [1]{%
 \ifnum #1\expandafter \@firstoftwo
 \else \expandafter \@secondoftwo
 \fi
}%
\providecommand \@ifx [1]{%
 \ifx #1\expandafter \@firstoftwo
 \else \expandafter \@secondoftwo
 \fi
}%
\providecommand \natexlab [1]{#1}%
\providecommand \enquote  [1]{``#1''}%
\providecommand \bibnamefont  [1]{#1}%
\providecommand \bibfnamefont [1]{#1}%
\providecommand \citenamefont [1]{#1}%
\providecommand \href@noop [0]{\@secondoftwo}%
\providecommand \href [0]{\begingroup \@sanitize@url \@href}%
\providecommand \@href[1]{\@@startlink{#1}\@@href}%
\providecommand \@@href[1]{\endgroup#1\@@endlink}%
\providecommand \@sanitize@url [0]{\catcode `\\12\catcode `\$12\catcode
  `\&12\catcode `\#12\catcode `\^12\catcode `\_12\catcode `\%12\relax}%
\providecommand \@@startlink[1]{}%
\providecommand \@@endlink[0]{}%
\providecommand \url  [0]{\begingroup\@sanitize@url \@url }%
\providecommand \@url [1]{\endgroup\@href {#1}{\urlprefix }}%
\providecommand \urlprefix  [0]{URL }%
\providecommand \Eprint [0]{\href }%
\providecommand \doibase [0]{http://dx.doi.org/}%
\providecommand \selectlanguage [0]{\@gobble}%
\providecommand \bibinfo  [0]{\@secondoftwo}%
\providecommand \bibfield  [0]{\@secondoftwo}%
\providecommand \translation [1]{[#1]}%
\providecommand \BibitemOpen [0]{}%
\providecommand \bibitemStop [0]{}%
\providecommand \bibitemNoStop [0]{.\EOS\space}%
\providecommand \EOS [0]{\spacefactor3000\relax}%
\providecommand \BibitemShut  [1]{\csname bibitem#1\endcsname}%
\let\auto@bib@innerbib\@empty
%</preamble>
\bibitem [{\citenamefont {Henkel}\ \emph {et~al.}(2008)\citenamefont {Henkel},
  \citenamefont {Hinrichsen},\ and\ \citenamefont {L\"{u}beck}}]{Henkel2008}%
  \BibitemOpen
  \bibfield  {author} {\bibinfo {author} {\bibfnamefont {M.}~\bibnamefont
  {Henkel}}, \bibinfo {author} {\bibfnamefont {H.}~\bibnamefont {Hinrichsen}},
  \ and\ \bibinfo {author} {\bibfnamefont {S.}~\bibnamefont {L\"{u}beck}},\
  }\href {\doibase 10.1007/978-1-4020-8765-3} {\emph {\bibinfo {title}
  {Non-Equilibrium Phase Transitions}}}\ (\bibinfo  {publisher} {Springer
  Netherlands},\ \bibinfo {year} {2008})\BibitemShut {NoStop}%
\bibitem [{\citenamefont {Henkel}\ and\ \citenamefont
  {Pleimling}(2010)}]{Henkel2010}%
  \BibitemOpen
  \bibfield  {author} {\bibinfo {author} {\bibfnamefont {M.}~\bibnamefont
  {Henkel}}\ and\ \bibinfo {author} {\bibfnamefont {M.}~\bibnamefont
  {Pleimling}},\ }\href {\doibase 10.1007/978-90-481-2869-3} {\emph {\bibinfo
  {title} {Non-Equilibrium Phase Transitions}}}\ (\bibinfo  {publisher}
  {Springer Netherlands},\ \bibinfo {year} {2010})\BibitemShut {NoStop}%
\bibitem [{\citenamefont {Domany}\ and\ \citenamefont
  {Kinzel}(1984)}]{Domany1984}%
  \BibitemOpen
  \bibfield  {author} {\bibinfo {author} {\bibfnamefont {E.}~\bibnamefont
  {Domany}}\ and\ \bibinfo {author} {\bibfnamefont {W.}~\bibnamefont
  {Kinzel}},\ }\href {\doibase 10.1103/PhysRevLett.53.311} {\bibfield
  {journal} {\bibinfo  {journal} {Phys. Rev. Lett.}\ }\textbf {\bibinfo
  {volume} {53}},\ \bibinfo {pages} {311} (\bibinfo {year} {1984})}\BibitemShut
  {NoStop}%
\bibitem [{\citenamefont {Bagnoli}\ \emph {et~al.}(2001)\citenamefont
  {Bagnoli}, \citenamefont {Boccara},\ and\ \citenamefont
  {Rechtman}}]{Bagnoli2001}%
  \BibitemOpen
  \bibfield  {author} {\bibinfo {author} {\bibfnamefont {F.}~\bibnamefont
  {Bagnoli}}, \bibinfo {author} {\bibfnamefont {N.}~\bibnamefont {Boccara}}, \
  and\ \bibinfo {author} {\bibfnamefont {R.}~\bibnamefont {Rechtman}},\ }\href
  {\doibase 10.1103/PhysRevE.63.046116} {\bibfield  {journal} {\bibinfo
  {journal} {Phys. Rev. E}\ }\textbf {\bibinfo {volume} {63}},\ \bibinfo
  {pages} {046116} (\bibinfo {year} {2001})}\BibitemShut {NoStop}%
\bibitem [{\citenamefont {Bagnoli}\ and\ \citenamefont
  {Rechtman}(2014)}]{Bagnoli2014}%
  \BibitemOpen
  \bibfield  {author} {\bibinfo {author} {\bibfnamefont {F.}~\bibnamefont
  {Bagnoli}}\ and\ \bibinfo {author} {\bibfnamefont {R.}~\bibnamefont
  {Rechtman}},\ }\href@noop {} {\bibfield  {journal} {\bibinfo  {journal}
  {arXiv preprint}\ } (\bibinfo {year} {2014})},\ \Eprint
  {http://arxiv.org/abs/1409.4284} {arXiv:1409.4284} \BibitemShut {NoStop}%
\bibitem [{\citenamefont {Hinrichsen}(2000)}]{Hinrichsen2000}%
  \BibitemOpen
  \bibfield  {author} {\bibinfo {author} {\bibfnamefont {H.}~\bibnamefont
  {Hinrichsen}},\ }\href {\doibase 10.1080/00018730050198152} {\bibfield
  {journal} {\bibinfo  {journal} {Adv. Phys.}\ }\textbf {\bibinfo {volume}
  {49}},\ \bibinfo {pages} {815} (\bibinfo {year} {2000})}\BibitemShut
  {NoStop}%
\bibitem [{\citenamefont {L\"{u}beck}(2004)}]{Lubeck2005}%
  \BibitemOpen
  \bibfield  {author} {\bibinfo {author} {\bibfnamefont {S.}~\bibnamefont
  {L\"{u}beck}},\ }\href@noop {} {\bibfield  {journal} {\bibinfo  {journal}
  {Int. J. Mod. Phys. B}\ }\textbf {\bibinfo {volume} {18}},\ \bibinfo {pages}
  {3977} (\bibinfo {year} {2004})}\BibitemShut {NoStop}%
\bibitem [{\citenamefont {Lesanovsky}\ \emph {et~al.}(2019)\citenamefont
  {Lesanovsky}, \citenamefont {Macieszczak},\ and\ \citenamefont
  {Garrahan}}]{Lesanovsky2019}%
  \BibitemOpen
  \bibfield  {author} {\bibinfo {author} {\bibfnamefont {I.}~\bibnamefont
  {Lesanovsky}}, \bibinfo {author} {\bibfnamefont {K.}~\bibnamefont
  {Macieszczak}}, \ and\ \bibinfo {author} {\bibfnamefont {J.~P.}\ \bibnamefont
  {Garrahan}},\ }\href@noop {} {\bibfield  {journal} {\bibinfo  {journal}
  {Quantum Sci. Technol.}\ }\textbf {\bibinfo {volume} {4}},\ \bibinfo {pages}
  {02LT02} (\bibinfo {year} {2019})}\BibitemShut {NoStop}%
\bibitem [{\citenamefont {Gillman}\ \emph {et~al.}(2020)\citenamefont
  {Gillman}, \citenamefont {Carollo},\ and\ \citenamefont
  {Lesanovsky}}]{Gillman2020}%
  \BibitemOpen
  \bibfield  {author} {\bibinfo {author} {\bibfnamefont {E.}~\bibnamefont
  {Gillman}}, \bibinfo {author} {\bibfnamefont {F.}~\bibnamefont {Carollo}}, \
  and\ \bibinfo {author} {\bibfnamefont {I.}~\bibnamefont {Lesanovsky}},\
  }\href {\doibase 10.1103/PhysRevLett.125.100403} {\bibfield  {journal}
  {\bibinfo  {journal} {Phys. Rev. Lett.}\ }\textbf {\bibinfo {volume} {125}},\
  \bibinfo {pages} {100403} (\bibinfo {year} {2020})}\BibitemShut {NoStop}%
\bibitem [{\citenamefont {Gillman}\ \emph
  {et~al.}(2021{\natexlab{a}})\citenamefont {Gillman}, \citenamefont
  {Carollo},\ and\ \citenamefont {Lesanovsky}}]{Gillman2021}%
  \BibitemOpen
  \bibfield  {author} {\bibinfo {author} {\bibfnamefont {E.}~\bibnamefont
  {Gillman}}, \bibinfo {author} {\bibfnamefont {F.}~\bibnamefont {Carollo}}, \
  and\ \bibinfo {author} {\bibfnamefont {I.}~\bibnamefont {Lesanovsky}},\
  }\href {\doibase 10.1103/PhysRevA.103.L040201} {\bibfield  {journal}
  {\bibinfo  {journal} {Phys. Rev. A}\ }\textbf {\bibinfo {volume} {103}},\
  \bibinfo {pages} {L040201} (\bibinfo {year}
  {2021}{\natexlab{a}})}\BibitemShut {NoStop}%
\bibitem [{\citenamefont {Gillman}\ \emph
  {et~al.}(2021{\natexlab{b}})\citenamefont {Gillman}, \citenamefont
  {Carollo},\ and\ \citenamefont {Lesanovsky}}]{Gillman2021b}%
  \BibitemOpen
  \bibfield  {author} {\bibinfo {author} {\bibfnamefont {E.}~\bibnamefont
  {Gillman}}, \bibinfo {author} {\bibfnamefont {F.}~\bibnamefont {Carollo}}, \
  and\ \bibinfo {author} {\bibfnamefont {I.}~\bibnamefont {Lesanovsky}},\
  }\href {\doibase 10.1103/PhysRevLett.127.230502} {\bibfield  {journal}
  {\bibinfo  {journal} {Phys. Rev. Lett.}\ }\textbf {\bibinfo {volume} {127}},\
  \bibinfo {pages} {230502} (\bibinfo {year} {2021}{\natexlab{b}})}\BibitemShut
  {NoStop}%
\bibitem [{\citenamefont {Zeiher}\ \emph {et~al.}(2016)\citenamefont {Zeiher},
  \citenamefont {Van~Bijnen}, \citenamefont {Schau{\ss}}, \citenamefont {Hild},
  \citenamefont {Choi}, \citenamefont {Pohl}, \citenamefont {Bloch},\ and\
  \citenamefont {Gross}}]{zeiher2016}%
  \BibitemOpen
  \bibfield  {author} {\bibinfo {author} {\bibfnamefont {J.}~\bibnamefont
  {Zeiher}}, \bibinfo {author} {\bibfnamefont {R.}~\bibnamefont {Van~Bijnen}},
  \bibinfo {author} {\bibfnamefont {P.}~\bibnamefont {Schau{\ss}}}, \bibinfo
  {author} {\bibfnamefont {S.}~\bibnamefont {Hild}}, \bibinfo {author}
  {\bibfnamefont {J.-y.}\ \bibnamefont {Choi}}, \bibinfo {author}
  {\bibfnamefont {T.}~\bibnamefont {Pohl}}, \bibinfo {author} {\bibfnamefont
  {I.}~\bibnamefont {Bloch}}, \ and\ \bibinfo {author} {\bibfnamefont
  {C.}~\bibnamefont {Gross}},\ }\href@noop {} {\bibfield  {journal} {\bibinfo
  {journal} {Nat, Phys.}\ }\textbf {\bibinfo {volume} {12}},\ \bibinfo {pages}
  {1095} (\bibinfo {year} {2016})}\BibitemShut {NoStop}%
\bibitem [{\citenamefont {Kim}\ \emph {et~al.}(2018)\citenamefont {Kim},
  \citenamefont {Park}, \citenamefont {Kim}, \citenamefont {Sim},\ and\
  \citenamefont {Ahn}}]{kim2018}%
  \BibitemOpen
  \bibfield  {author} {\bibinfo {author} {\bibfnamefont {H.}~\bibnamefont
  {Kim}}, \bibinfo {author} {\bibfnamefont {Y.}~\bibnamefont {Park}}, \bibinfo
  {author} {\bibfnamefont {K.}~\bibnamefont {Kim}}, \bibinfo {author}
  {\bibfnamefont {H.-S.}\ \bibnamefont {Sim}}, \ and\ \bibinfo {author}
  {\bibfnamefont {J.}~\bibnamefont {Ahn}},\ }\href {\doibase
  10.1103/PhysRevLett.120.180502} {\bibfield  {journal} {\bibinfo  {journal}
  {Phys. Rev. Lett.}\ }\textbf {\bibinfo {volume} {120}},\ \bibinfo {pages}
  {180502} (\bibinfo {year} {2018})}\BibitemShut {NoStop}%
\bibitem [{\citenamefont {Browaeys}\ and\ \citenamefont
  {Lahaye}(2020)}]{browaeys2020}%
  \BibitemOpen
  \bibfield  {author} {\bibinfo {author} {\bibfnamefont {A.}~\bibnamefont
  {Browaeys}}\ and\ \bibinfo {author} {\bibfnamefont {T.}~\bibnamefont
  {Lahaye}},\ }\href@noop {} {\bibfield  {journal} {\bibinfo  {journal} {Nat.
  Phys.}\ }\textbf {\bibinfo {volume} {16}},\ \bibinfo {pages} {132} (\bibinfo
  {year} {2020})}\BibitemShut {NoStop}%
\bibitem [{\citenamefont {Ebadi}\ \emph {et~al.}(2020)\citenamefont {Ebadi},
  \citenamefont {Wang}, \citenamefont {Levine}, \citenamefont {Keesling},
  \citenamefont {Semeghini}, \citenamefont {Omran}, \citenamefont {Bluvstein},
  \citenamefont {Samajdar}, \citenamefont {Pichler}, \citenamefont {Ho} \emph
  {et~al.}}]{ebadi2020}%
  \BibitemOpen
  \bibfield  {author} {\bibinfo {author} {\bibfnamefont {S.}~\bibnamefont
  {Ebadi}}, \bibinfo {author} {\bibfnamefont {T.~T.}\ \bibnamefont {Wang}},
  \bibinfo {author} {\bibfnamefont {H.}~\bibnamefont {Levine}}, \bibinfo
  {author} {\bibfnamefont {A.}~\bibnamefont {Keesling}}, \bibinfo {author}
  {\bibfnamefont {G.}~\bibnamefont {Semeghini}}, \bibinfo {author}
  {\bibfnamefont {A.}~\bibnamefont {Omran}}, \bibinfo {author} {\bibfnamefont
  {D.}~\bibnamefont {Bluvstein}}, \bibinfo {author} {\bibfnamefont
  {R.}~\bibnamefont {Samajdar}}, \bibinfo {author} {\bibfnamefont
  {H.}~\bibnamefont {Pichler}}, \bibinfo {author} {\bibfnamefont {W.~W.}\
  \bibnamefont {Ho}},  \emph {et~al.},\ }\href@noop {} {\bibfield  {journal}
  {\bibinfo  {journal} {arXiv:2012.12281}\ } (\bibinfo {year}
  {2020})}\BibitemShut {NoStop}%
\bibitem [{\citenamefont {Wiesner}(2009)}]{Wiesner2009}%
  \BibitemOpen
  \bibfield  {author} {\bibinfo {author} {\bibfnamefont {K.}~\bibnamefont
  {Wiesner}},\ }\enquote {\bibinfo {title} {Quantum cellular automata},}\ in\
  \href {\doibase 10.1007/978-0-387-30440-3_426} {\emph {\bibinfo {booktitle}
  {Encyclopedia of Complexity and Systems Science}}},\ \bibinfo {editor}
  {edited by\ \bibinfo {editor} {\bibfnamefont {R.~A.}\ \bibnamefont
  {Meyers}}}\ (\bibinfo  {publisher} {Springer New York},\ \bibinfo {address}
  {New York, NY},\ \bibinfo {year} {2009})\ pp.\ \bibinfo {pages}
  {7154--7164}\BibitemShut {NoStop}%
\bibitem [{\citenamefont {Cirac}\ \emph {et~al.}(2017)\citenamefont {Cirac},
  \citenamefont {Perez-Garcia}, \citenamefont {Schuch},\ and\ \citenamefont
  {Verstraete}}]{Cirac2017}%
  \BibitemOpen
  \bibfield  {author} {\bibinfo {author} {\bibfnamefont {J.~I.}\ \bibnamefont
  {Cirac}}, \bibinfo {author} {\bibfnamefont {D.}~\bibnamefont {Perez-Garcia}},
  \bibinfo {author} {\bibfnamefont {N.}~\bibnamefont {Schuch}}, \ and\ \bibinfo
  {author} {\bibfnamefont {F.}~\bibnamefont {Verstraete}},\ }\href {\doibase
  10.1088/1742-5468/aa7e55} {\bibfield  {journal} {\bibinfo  {journal} {J.
  Stat. Mech.-Theory E.}\ }\textbf {\bibinfo {volume} {2017}},\ \bibinfo
  {pages} {083105} (\bibinfo {year} {2017})}\BibitemShut {NoStop}%
\bibitem [{\citenamefont {Arrighi}(2019)}]{Arrighi2019}%
  \BibitemOpen
  \bibfield  {author} {\bibinfo {author} {\bibfnamefont {P.}~\bibnamefont
  {Arrighi}},\ }\href {\doibase 10.1007/s11047-019-09762-6} {\bibfield
  {journal} {\bibinfo  {journal} {Nat. Comput.}\ }\textbf {\bibinfo {volume}
  {18}},\ \bibinfo {pages} {885} (\bibinfo {year} {2019})}\BibitemShut
  {NoStop}%
\bibitem [{\citenamefont {Farrelly}(2020)}]{Farrelly2020}%
  \BibitemOpen
  \bibfield  {author} {\bibinfo {author} {\bibfnamefont {T.}~\bibnamefont
  {Farrelly}},\ }\href {\doibase 10.22331/q-2020-11-30-368} {\bibfield
  {journal} {\bibinfo  {journal} {{Quantum}}\ }\textbf {\bibinfo {volume}
  {4}},\ \bibinfo {pages} {368} (\bibinfo {year} {2020})}\BibitemShut {NoStop}%
\bibitem [{\citenamefont {Hillberry}\ \emph {et~al.}(2021)\citenamefont
  {Hillberry}, \citenamefont {Jones}, \citenamefont {Vargas}, \citenamefont
  {Rall}, \citenamefont {Halpern}, \citenamefont {Bao}, \citenamefont
  {Notarnicola}, \citenamefont {Montangero},\ and\ \citenamefont
  {Carr}}]{Hillberry2021}%
  \BibitemOpen
  \bibfield  {author} {\bibinfo {author} {\bibfnamefont {L.~E.}\ \bibnamefont
  {Hillberry}}, \bibinfo {author} {\bibfnamefont {M.~T.}\ \bibnamefont
  {Jones}}, \bibinfo {author} {\bibfnamefont {D.~L.}\ \bibnamefont {Vargas}},
  \bibinfo {author} {\bibfnamefont {P.}~\bibnamefont {Rall}}, \bibinfo {author}
  {\bibfnamefont {N.~Y.}\ \bibnamefont {Halpern}}, \bibinfo {author}
  {\bibfnamefont {N.}~\bibnamefont {Bao}}, \bibinfo {author} {\bibfnamefont
  {S.}~\bibnamefont {Notarnicola}}, \bibinfo {author} {\bibfnamefont
  {S.}~\bibnamefont {Montangero}}, \ and\ \bibinfo {author} {\bibfnamefont
  {L.~D.}\ \bibnamefont {Carr}},\ }\href {\doibase 10.1088/2058-9565/ac1c41}
  {\bibfield  {journal} {\bibinfo  {journal} {Quantum Sci. Technol.}\ }\textbf
  {\bibinfo {volume} {6}},\ \bibinfo {pages} {045017} (\bibinfo {year}
  {2021})}\BibitemShut {NoStop}%
\bibitem [{\citenamefont {Beer}\ \emph {et~al.}(2020)\citenamefont {Beer},
  \citenamefont {Bondarenko}, \citenamefont {Farrelly}, \citenamefont
  {Osborne}, \citenamefont {Salzmann}, \citenamefont {Scheiermann},\ and\
  \citenamefont {Wolf}}]{Beer2020}%
  \BibitemOpen
  \bibfield  {author} {\bibinfo {author} {\bibfnamefont {K.}~\bibnamefont
  {Beer}}, \bibinfo {author} {\bibfnamefont {D.}~\bibnamefont {Bondarenko}},
  \bibinfo {author} {\bibfnamefont {T.}~\bibnamefont {Farrelly}}, \bibinfo
  {author} {\bibfnamefont {T.~J.}\ \bibnamefont {Osborne}}, \bibinfo {author}
  {\bibfnamefont {R.}~\bibnamefont {Salzmann}}, \bibinfo {author}
  {\bibfnamefont {D.}~\bibnamefont {Scheiermann}}, \ and\ \bibinfo {author}
  {\bibfnamefont {R.}~\bibnamefont {Wolf}},\ }\href@noop {} {\bibfield
  {journal} {\bibinfo  {journal} {Nat. Commun.}\ }\textbf {\bibinfo {volume}
  {11}},\ \bibinfo {pages} {808} (\bibinfo {year} {2020})}\BibitemShut
  {NoStop}%
\bibitem [{\citenamefont {Bour{\'e}}\ \emph {et~al.}(2012)\citenamefont
  {Bour{\'e}}, \citenamefont {Fat{\`e}s},\ and\ \citenamefont
  {Chevrier}}]{Boure2012}%
  \BibitemOpen
  \bibfield  {author} {\bibinfo {author} {\bibfnamefont {O.}~\bibnamefont
  {Bour{\'e}}}, \bibinfo {author} {\bibfnamefont {N.}~\bibnamefont
  {Fat{\`e}s}}, \ and\ \bibinfo {author} {\bibfnamefont {V.}~\bibnamefont
  {Chevrier}},\ }\href {\doibase 10.1007/s11047-012-9340-y} {\bibfield
  {journal} {\bibinfo  {journal} {Nat. Comput.}\ }\textbf {\bibinfo {volume}
  {11}},\ \bibinfo {pages} {553} (\bibinfo {year} {2012})}\BibitemShut
  {NoStop}%
\bibitem [{\citenamefont {Bandini}\ \emph {et~al.}(2012)\citenamefont
  {Bandini}, \citenamefont {Bonomi},\ and\ \citenamefont
  {Vizzari}}]{Bandini2012}%
  \BibitemOpen
  \bibfield  {author} {\bibinfo {author} {\bibfnamefont {S.}~\bibnamefont
  {Bandini}}, \bibinfo {author} {\bibfnamefont {A.}~\bibnamefont {Bonomi}}, \
  and\ \bibinfo {author} {\bibfnamefont {G.}~\bibnamefont {Vizzari}},\ }\href
  {\doibase 10.1007/s11047-012-9310-4} {\bibfield  {journal} {\bibinfo
  {journal} {Nat. Comput.}\ }\textbf {\bibinfo {volume} {11}},\ \bibinfo
  {pages} {277} (\bibinfo {year} {2012})}\BibitemShut {NoStop}%
\bibitem [{\citenamefont {Fat{\`e}s}(2013)}]{Fates2013}%
  \BibitemOpen
  \bibfield  {author} {\bibinfo {author} {\bibfnamefont {N.}~\bibnamefont
  {Fat{\`e}s}},\ }in\ \href@noop {} {\emph {\bibinfo {booktitle} {Cellular
  Automata and Discrete Complex Systems}}},\ \bibinfo {editor} {edited by\
  \bibinfo {editor} {\bibfnamefont {J.}~\bibnamefont {Kari}}, \bibinfo {editor}
  {\bibfnamefont {M.}~\bibnamefont {Kutrib}}, \ and\ \bibinfo {editor}
  {\bibfnamefont {A.}~\bibnamefont {Malcher}}}\ (\bibinfo  {publisher}
  {Springer Berlin Heidelberg},\ \bibinfo {address} {Berlin, Heidelberg},\
  \bibinfo {year} {2013})\ pp.\ \bibinfo {pages} {15--30}\BibitemShut {NoStop}%
\bibitem [{\citenamefont {Harris}(1974)}]{Harris1974}%
  \BibitemOpen
  \bibfield  {author} {\bibinfo {author} {\bibfnamefont {T.~E.}\ \bibnamefont
  {Harris}},\ }\href {\doibase 10.1214/aop/1176996493} {\bibfield  {journal}
  {\bibinfo  {journal} {The Annals of Probability}\ }\textbf {\bibinfo {volume}
  {2}},\ \bibinfo {pages} {969 } (\bibinfo {year} {1974})}\BibitemShut
  {NoStop}%
\bibitem [{\citenamefont {Marcuzzi}\ \emph {et~al.}(2016)\citenamefont
  {Marcuzzi}, \citenamefont {Buchhold}, \citenamefont {Diehl},\ and\
  \citenamefont {Lesanovsky}}]{Marcuzzi2016}%
  \BibitemOpen
  \bibfield  {author} {\bibinfo {author} {\bibfnamefont {M.}~\bibnamefont
  {Marcuzzi}}, \bibinfo {author} {\bibfnamefont {M.}~\bibnamefont {Buchhold}},
  \bibinfo {author} {\bibfnamefont {S.}~\bibnamefont {Diehl}}, \ and\ \bibinfo
  {author} {\bibfnamefont {I.}~\bibnamefont {Lesanovsky}},\ }\href {\doibase
  10.1103/PhysRevLett.116.245701} {\bibfield  {journal} {\bibinfo  {journal}
  {Phys. Rev. Lett.}\ }\textbf {\bibinfo {volume} {116}},\ \bibinfo {pages}
  {245701} (\bibinfo {year} {2016})}\BibitemShut {NoStop}%
\bibitem [{\citenamefont {Buchhold}\ \emph {et~al.}(2017)\citenamefont
  {Buchhold}, \citenamefont {Everest}, \citenamefont {Marcuzzi}, \citenamefont
  {Lesanovsky},\ and\ \citenamefont {Diehl}}]{Buchhold2017}%
  \BibitemOpen
  \bibfield  {author} {\bibinfo {author} {\bibfnamefont {M.}~\bibnamefont
  {Buchhold}}, \bibinfo {author} {\bibfnamefont {B.}~\bibnamefont {Everest}},
  \bibinfo {author} {\bibfnamefont {M.}~\bibnamefont {Marcuzzi}}, \bibinfo
  {author} {\bibfnamefont {I.}~\bibnamefont {Lesanovsky}}, \ and\ \bibinfo
  {author} {\bibfnamefont {S.}~\bibnamefont {Diehl}},\ }\href {\doibase
  10.1103/PhysRevB.95.014308} {\bibfield  {journal} {\bibinfo  {journal} {Phys.
  Rev. B}\ }\textbf {\bibinfo {volume} {95}},\ \bibinfo {pages} {014308}
  (\bibinfo {year} {2017})}\BibitemShut {NoStop}%
\bibitem [{\citenamefont {Roscher}\ \emph {et~al.}(2018)\citenamefont
  {Roscher}, \citenamefont {Diehl},\ and\ \citenamefont
  {Buchhold}}]{Roscher2018}%
  \BibitemOpen
  \bibfield  {author} {\bibinfo {author} {\bibfnamefont {D.}~\bibnamefont
  {Roscher}}, \bibinfo {author} {\bibfnamefont {S.}~\bibnamefont {Diehl}}, \
  and\ \bibinfo {author} {\bibfnamefont {M.}~\bibnamefont {Buchhold}},\ }\href
  {\doibase 10.1103/PhysRevA.98.062117} {\bibfield  {journal} {\bibinfo
  {journal} {Phys. Rev. A}\ }\textbf {\bibinfo {volume} {98}},\ \bibinfo
  {pages} {062117} (\bibinfo {year} {2018})}\BibitemShut {NoStop}%
\bibitem [{\citenamefont {Carollo}\ \emph {et~al.}(2019)\citenamefont
  {Carollo}, \citenamefont {Gillman}, \citenamefont {Weimer},\ and\
  \citenamefont {Lesanovsky}}]{Carollo2019}%
  \BibitemOpen
  \bibfield  {author} {\bibinfo {author} {\bibfnamefont {F.}~\bibnamefont
  {Carollo}}, \bibinfo {author} {\bibfnamefont {E.}~\bibnamefont {Gillman}},
  \bibinfo {author} {\bibfnamefont {H.}~\bibnamefont {Weimer}}, \ and\ \bibinfo
  {author} {\bibfnamefont {I.}~\bibnamefont {Lesanovsky}},\ }\href {\doibase
  10.1103/PhysRevLett.123.100604} {\bibfield  {journal} {\bibinfo  {journal}
  {Phys. Rev. Lett.}\ }\textbf {\bibinfo {volume} {123}},\ \bibinfo {pages}
  {100604} (\bibinfo {year} {2019})}\BibitemShut {NoStop}%
\bibitem [{\citenamefont {Gillman}\ \emph {et~al.}(2019)\citenamefont
  {Gillman}, \citenamefont {Carollo},\ and\ \citenamefont
  {Lesanovsky}}]{Gillman2019}%
  \BibitemOpen
  \bibfield  {author} {\bibinfo {author} {\bibfnamefont {E.}~\bibnamefont
  {Gillman}}, \bibinfo {author} {\bibfnamefont {F.}~\bibnamefont {Carollo}}, \
  and\ \bibinfo {author} {\bibfnamefont {I.}~\bibnamefont {Lesanovsky}},\
  }\href {\doibase 10.1088/1367-2630/ab43b0} {\bibfield  {journal} {\bibinfo
  {journal} {New J. Phys.}\ }\textbf {\bibinfo {volume} {21}},\ \bibinfo
  {pages} {093064} (\bibinfo {year} {2019})}\BibitemShut {NoStop}%
\bibitem [{\citenamefont {Jo}\ \emph {et~al.}(2021)\citenamefont {Jo},
  \citenamefont {Lee}, \citenamefont {Choi},\ and\ \citenamefont
  {Kahng}}]{Jo2021}%
  \BibitemOpen
  \bibfield  {author} {\bibinfo {author} {\bibfnamefont {M.}~\bibnamefont
  {Jo}}, \bibinfo {author} {\bibfnamefont {J.}~\bibnamefont {Lee}}, \bibinfo
  {author} {\bibfnamefont {K.}~\bibnamefont {Choi}}, \ and\ \bibinfo {author}
  {\bibfnamefont {B.}~\bibnamefont {Kahng}},\ }\href {\doibase
  10.1103/PhysRevResearch.3.013238} {\bibfield  {journal} {\bibinfo  {journal}
  {Phys. Rev. Research}\ }\textbf {\bibinfo {volume} {3}},\ \bibinfo {pages}
  {013238} (\bibinfo {year} {2021})}\BibitemShut {NoStop}%
\bibitem [{\citenamefont {Wolfram}(1983)}]{Wolfram1983}%
  \BibitemOpen
  \bibfield  {author} {\bibinfo {author} {\bibfnamefont {S.}~\bibnamefont
  {Wolfram}},\ }\href {\doibase 10.1103/RevModPhys.55.601} {\bibfield
  {journal} {\bibinfo  {journal} {Rev. Mod. Phys.}\ }\textbf {\bibinfo {volume}
  {55}},\ \bibinfo {pages} {601} (\bibinfo {year} {1983})}\BibitemShut
  {NoStop}%
\bibitem [{\citenamefont {Wolfram}(2002)}]{Wolfram2002}%
  \BibitemOpen
  \bibfield  {author} {\bibinfo {author} {\bibfnamefont {S.}~\bibnamefont
  {Wolfram}},\ }\href {https://www.wolframscience.com} {\emph {\bibinfo {title}
  {A New Kind of Science}}}\ (\bibinfo  {publisher} {Wolfram Media},\ \bibinfo
  {year} {2002})\BibitemShut {NoStop}%
\bibitem [{SM()}]{SM}%
  \BibitemOpen
  \href@noop {} {\bibinfo  {journal} {see Supplemental Material for details}\
  }\BibitemShut {NoStop}%
\bibitem [{\citenamefont {Jo}\ \emph {et~al.}(2019)\citenamefont {Jo},
  \citenamefont {Um},\ and\ \citenamefont {Kahng}}]{Jo2019}%
  \BibitemOpen
\bibfield  {journal} {  }\bibfield  {author} {\bibinfo {author} {\bibfnamefont
  {M.}~\bibnamefont {Jo}}, \bibinfo {author} {\bibfnamefont {J.}~\bibnamefont
  {Um}}, \ and\ \bibinfo {author} {\bibfnamefont {B.}~\bibnamefont {Kahng}},\
  }\href {\doibase 10.1103/PhysRevE.99.032131} {\bibfield  {journal} {\bibinfo
  {journal} {Phys. Rev. E}\ }\textbf {\bibinfo {volume} {99}},\ \bibinfo
  {pages} {032131} (\bibinfo {year} {2019})}\BibitemShut {NoStop}%
\bibitem [{\citenamefont {Jo}\ and\ \citenamefont {Kahng}(2020)}]{Jo2020}%
  \BibitemOpen
  \bibfield  {author} {\bibinfo {author} {\bibfnamefont {M.}~\bibnamefont
  {Jo}}\ and\ \bibinfo {author} {\bibfnamefont {B.}~\bibnamefont {Kahng}},\
  }\href {\doibase 10.1103/PhysRevE.101.022121} {\bibfield  {journal} {\bibinfo
   {journal} {Phys. Rev. E}\ }\textbf {\bibinfo {volume} {101}},\ \bibinfo
  {pages} {022121} (\bibinfo {year} {2020})}\BibitemShut {NoStop}%
\bibitem [{\citenamefont {Nigmatullin}\ \emph {et~al.}(2021)\citenamefont
  {Nigmatullin}, \citenamefont {Wagner},\ and\ \citenamefont
  {Brennen}}]{Nigmatullin2021}%
  \BibitemOpen
  \bibfield  {author} {\bibinfo {author} {\bibfnamefont {R.}~\bibnamefont
  {Nigmatullin}}, \bibinfo {author} {\bibfnamefont {E.}~\bibnamefont {Wagner}},
  \ and\ \bibinfo {author} {\bibfnamefont {G.~K.}\ \bibnamefont {Brennen}},\
  }\href {\doibase 10.1103/PhysRevResearch.3.043167} {\bibfield  {journal}
  {\bibinfo  {journal} {Phys. Rev. Research}\ }\textbf {\bibinfo {volume}
  {3}},\ \bibinfo {pages} {043167} (\bibinfo {year} {2021})}\BibitemShut
  {NoStop}%
\end{thebibliography}%

% SUPP MATERIAL
\onecolumngrid
\newpage

\pagebreak
\widetext

\begin{center}
\textbf{\large Supplemental Material}
\end{center}

\setcounter{section}{0}
\setcounter{equation}{0}
\setcounter{figure}{0}
\setcounter{table}{0}
\setcounter{page}{1}
\makeatletter

\renewcommand\thesection{S\arabic{section}}
\renewcommand{\theequation}{S\arabic{equation}}
\renewcommand{\thefigure}{S\arabic{figure}}
\renewcommand{\thetable}{S\arabic{table}}
\renewcommand{\bibnumfmt}[1]{[S#1]}

\section{Mean-Field Equations for the $(1+1)D$ QCA in the main text}
In this section we give details on our mean-field analysis for the QCA model discussed in the main text. We directly present results for the asynchronous gate, since the synchronous discrete-time evolution can be obtained by setting the asynchronicity parameter to zero. 

In our mean-field analysis, we assume that the reduced state of single rows of the lattice is a product state at any time. This amounts to neglecting correlations within single rows. We furthermore assume that this reduced state is homogeneous, i.e. local single-site properties are equal for all sites belonging to a same row. Because of this assumption, we can drop the index $k$ in the notation of this section.  In particular, single-site properties of the reduced state of the QCA at time $t$ are described by the density matrix  
$$
\rho_t=\begin{pmatrix}
\langle n\rangle_t & \frac{x_t-iy_t}{2} \\
\frac{x_t+iy_t}{2} & \langle \bar{n}\rangle_t 
\end{pmatrix}\, ,
$$
where $x_t=\langle \sigma_x\rangle_t$ and $y_t=\langle \sigma_y\rangle_t$.

The task is then to find a discrete-time equation for updating such a single-site density matrix. Here, we define the update by considering how a single gate application modifies the target site. Due to our assumptions, control sites are in a product state with same single-site density matrix, while the target site before the update is in the empty state. We thus write the state of the three control sites and of the target one, before the update, as
\begin{equation}
\tilde{\rho}_t=\rho_{t}\otimes \rho_{t}\otimes \rho_{t}\otimes \bar{n} .
\label{SM:row-t}
\end{equation}
For later convenience, in the above expression the first two entries of the tensor product represent the two controls on the side of the target site, while the third entry represents the central control which is exactly above the target site [see also Fig.~\ref{Fig1}(a) in the main text]. The last entry, here consisting of a projector on the empty state, is instead the state of the target site. Exploiting the state $\tilde{\rho}_t$ and considering a single gate application, we can define, within our mean-field analysis, the single-site state of the row at time $t+1$ as 
\begin{equation}
    \rho_{t+1}=\Tr_{123}\left(G \tilde{\rho}_t G^\dagger \right)\, .
    \label{update-mf}
\end{equation}
In the above equation, $\Tr_{123}$ denotes the trace over the first three entries of the tensor product in Eq.~\eqref{SM:row-t}. The gate $G$ is the four-body gate implementing the local update rules considered in the main text. Here, we rewrite it as 
$$
G=\Pi\otimes A + \bar{\Pi}\otimes D\, , \qquad \mbox{ where }\qquad D=\sqrt{1-\lambda} B+\sqrt{\lambda}C
$$
and 
\begin{equation}
    A=n\otimes U_{\circ\bullet\circ}+\bar{n} \otimes \id\, ,\qquad B=n\otimes U_{\bullet} +\bar{n}\otimes U_{\circ}\, , \qquad C=\sigma_+ \otimes U_\bullet U_+ -\sigma_-\otimes U_{\circ}U_+^\dagger \, .
\end{equation}
The unitary operators appearing in the above equations  are the same reported in the main text. Moreover, we recall here that, in the notation of this section, $\Pi=\bar{n}\otimes \bar{n}$ acts only on the external control sites, and $\bar{n}=\ket{\circ}\!\bra{\circ}$. The other projector is instead $\bar{\Pi}=\id\otimes \id -\Pi$ and also acts solely on the above mentioned control sites. 

With the form of the gate and of the state $\tilde{\rho}_t$, we can proceed to evaluate the trace operation in Eq.~\eqref{update-mf}. We do this in two steps. First, we take the trace with respect to the external control sites, which are described by the first two entries of the tensor products. Defining $\rho^{12}_t=\rho_{t}\otimes \rho_{t}$ and $\rho^{34}_t=\rho_{t}\otimes \bar{n}$, so that $\tilde{\rho}_t=\rho^{12}\otimes \rho^{34} $, we find 
\begin{equation}
\Tr_{12}\left(G\tilde{\rho}_t G^\dagger\right)=\langle \Pi\rangle^{12}_t A\rho^{34}_tA^\dagger +\langle \bar{\Pi}\rangle^{12}_t  D\rho^{34}_tD^\dagger\, , 
\label{first-trace}
\end{equation}
where $\langle \cdot \rangle^{12}$ denotes expectation value with respect to the state $\rho^{12}$. 

Then, we can take the trace with respect to the central control site, which was represented by the third entry of the tensor product in Eq.~\eqref{SM:row-t}. We do this considering the two different terms of the above equation separately. We start with 
$$
\Tr_3\left(A\rho^{34}_tA^\dagger\right)=\langle n\rangle^3_t U_{\circ\bullet\circ}\bar{n}U_{\circ\bullet\circ}^\dagger + \langle \bar{n} \rangle^3_t \bar{n}\, ,
$$
where $\langle \cdot \rangle^{3}_t$ denotes expectation value with respect to the state of the central control site, which is --- as for the other control sites --- $\rho_t$. The second term on the right-hand-side of Eq.~\eqref{first-trace} is slightly more involved since $D$ is made of two terms. Indeed, we have 
$$
\Tr_{3}\left(D\rho^{34}_tD^\dagger\right)=(1-\lambda)\Tr_{3}\left(B\rho^{34}_tB^\dagger\right)+\lambda \Tr_{3}\left(C\rho^{34}_tC^\dagger\right)+\sqrt{\lambda}\sqrt{1-\lambda}\left[\Tr_{3}\left(B\rho^{34}_tC^\dagger\right)+\Tr_{3}\left(C\rho^{34}_tB^\dagger\right)\right]\, .
$$
A straightforward calculation gives 
\begin{equation*}
    \Tr_{3}\left(B\rho^{34}_tB^\dagger\right)=\langle n\rangle^3_t U_\bullet \bar{n}U_\bullet^\dagger +\langle \bar{n}\rangle^3_t U_{\circ}\bar{n}U_{\circ}^\dagger \, ,
\end{equation*}
\begin{equation*}
    \Tr_{3}\left(C\rho^{34}_tC^\dagger\right)=\langle n\rangle^3_t U_{\circ}U_+^\dagger\bar{n}U_+ U_{\circ}^\dagger +\langle \bar{n}\rangle^3_t U_{\bullet}U_+\bar{n}U_+^\dagger U_{\bullet}^\dagger \, ,
\end{equation*}

\begin{equation*}
    \Tr_{3}\left(B\rho^{34}_tC^\dagger\right)=\langle \sigma_-\rangle^3_t U_\bullet \bar{n}U_+^\dagger U_\bullet^\dagger -\langle \sigma_+\rangle^3_t U_{\circ}\bar{n}U_+U_{\circ}^\dagger 
\end{equation*}
as well as $\Tr_{3}\left(C\rho^{34}_tB^\dagger\right)=\Tr_{3}\left(B\rho^{34}_tC^\dagger\right)^\dagger$. Putting everything together, we arrive at 
\begin{equation}
\begin{split}
    \rho_{t+1}=&\langle \Pi\rangle_t \left(\langle n\rangle_t U_{\circ\bullet\circ}\bar{n}U_{\circ\bullet\circ}^\dagger +\langle \bar{n}\rangle_t \bar{n} \right)+\\
    &+(1-\lambda)\langle \Pi\rangle_t \left(\langle n\rangle_t U_{\bullet}\bar{n}U_{\bullet}^\dagger+\langle \bar{n}\rangle_t U_{\circ}\bar{n}U_{\circ }^\dagger\right)+\\
    &+\lambda \langle \Pi\rangle_t \left(\langle n\rangle_t U_{\circ} U_+^\dagger \bar{n} U_+  U_{\circ}^\dagger + \langle \bar{n}\rangle_t U_{\bullet} U_+ \bar{n} U_+^\dagger   U_{\bullet}^\dagger\right)+\\
    &+\sqrt{\lambda}\sqrt{1-\lambda} \langle \Pi \rangle_t \left(\langle \sigma_-\rangle_t U_\bullet\bar{n}U_+^\dagger U_\bullet^\dagger -\langle \sigma_+\rangle_t U_{\circ}\bar{n}U_+ U_{\circ}^\dagger + \mathrm{h.c.}\right)\, .
    \end{split}
\end{equation}
This iterative equation allows one to study the dynamics of the QCA as well as its stationary state properties within a mean-field investigation. For instance, to obtain the stationary phase diagram of our QCA reported in the main text, we have numerically simulated the above equation over a sufficiently large number of discrete time-steps. 

Additionally, with this expression, we can also compute iterative equations for the dynamics of expectation values. For instance, for the projector $n$ we find
\begin{equation}
\begin{split}
    \langle n\rangle_{t+1}=\Tr\left(\rho_{t+1}n\right)=&\langle \Pi\rangle_t \langle n\rangle_t q_{\circ\bullet\circ}+\langle \bar{\Pi}\rangle_t \langle n\rangle_t \left[(1-\lambda)p_\bullet+\lambda \left(\sqrt{p_{\circ}}\sqrt{q}+\sqrt{p}\sqrt{q_{\circ}}\right)^{2}  \right]+\\
    &+\langle \bar{\Pi}\rangle_t \langle \bar{n}\rangle_t \left[(1-\lambda) p_{\circ}+\lambda (\sqrt{p_\bullet}\sqrt{q}-\sqrt{p}\sqrt{q_\bullet})^{2}\right]+\\
    &+\sqrt{\lambda}\sqrt{1-\lambda}\langle \bar{\Pi}\rangle_t \langle \sigma^y\rangle_t\left[\sqrt{q}\left(p_\bullet+p_{\circ}\right)+\sqrt{p}\left(\sqrt{p_\circ}\sqrt{q_\circ}-\sqrt{p_\bullet}\sqrt{q_\bullet}\right)\right]\, .
    \end{split}
\label{eqn:full_qca_mfe_sm}
\end{equation}
The coefficients in the above equation are reported in the main text. We also note that for the synchronous case $\lambda=0$, one recovers the equation 

\begin{equation}
\begin{split}
    \langle n\rangle_{t+1}=\langle \Pi\rangle_t \langle n\rangle_t q_{\circ\bullet\circ}+\langle \bar{\Pi}\rangle_t \langle n\rangle_t p_\bullet + \langle \bar{\Pi}\rangle_t \langle \bar{n}\rangle_t p_{\circ}\, ,
    \end{split}
\label{eqn:ccp_qca_mfe}
\end{equation}
which is also reported in the main text.

\section{Mean-Field Equations for continuous-time Quantum and Classical Contact Process}

In this section, we briefly introduce a quantum contact process model, which, as discussed in the main text, is connected to the physics displayed by our $(1+1)$D QCA. 

The quantum contact process that we consider here consists of a one-dimensional spin-$1/2$ system which undergoes a continuous-time Markovian dynamics. In particular, the dynamics of any operator $X$ of the system obeys the so-called Heisenberg  equation \cite{Marcuzzi2016}
\begin{equation}
\begin{split}
\dot{X}_t= &i[H,X_t]+\gamma \sum_{k} \left(\sigma_k^+ X_t  \sigma_k^--\frac{1}{2}\left\{n_k,X_t\right\}\right)+\kappa_{\mathrm{c}}\sum_{k} \left(\bar{\Pi}_k\sigma_k^+ X_t  \sigma_k^-\bar{\Pi}_k-\frac{1}{2}\left\{n_k\bar{\Pi}_k,X_t\right\}\right)+\\
&+\kappa_{\mathrm{b}}\sum_{k} \left(\bar{\Pi}_k\sigma_k^- X_t  \sigma_k^+\bar{\Pi}_k-\frac{1}{2}\left\{\bar{n}_k\bar{\Pi}_k,X_t\right\}\right)\, .
\end{split}
\label{SM:QCP-gen}
\end{equation}
In the above equation, the first term accounts for the coherent dynamical contribution associated with the system Hamiltonian. On the other hand, the remaining terms describe incoherent probabilistic processes, which are in fact exactly those of the (classical) contact process. Indeed, the term proportional to $\gamma$ implements the decay process $\bullet \rightsquigarrow\circ$, the term proportional to $\kappa_{\mathrm{c}}$ is instead the coagulation process ($\bullet \bullet \rightsquigarrow \bullet \circ$), while the one proportional to $\kappa_{\mathrm{b}}$ is the branching process ($\circ \bullet \rightsquigarrow \bullet \bullet$). Note that, due to the presence of the projector $\bar{\Pi}_k$ these two latter processes can  occur at site $k$ only if at least an occupation is present in the neighboring sites of $k$. 

The Hamiltonian of the quantum contact process is the following 
$$
H=\Omega \sum_k\bar{\Pi}_k \sigma_k^x  ~.
$$
This Hamiltonian implements both a branching and a coagulation process at the same coherent rate, $\Omega$, via constrained Rabi oscillations at site $k$ that can only occur if the neighbors of $k$ are not simultaneously in the empty state.

We now want to obtain mean-field dynamical equations of motion for the above system. To this end, we consider single-site operators, compute their time derivative according to the action of the generator in Eq.~\eqref{SM:QCP-gen}, and make the assumption of uncorrelated state for the system. Following closely the calculations reported in Ref.~\cite{Buchhold2017}, the mean-field equations of motion for the number operator $n$ (also assuming a homogeneous state)  for the quantum contact process is given by
\begin{align}
\partial_{t} \braket{n}_t &= - \gamma \braket{n}_t+ \Omega\braket{\bar{\Pi}}_t \braket{\sigma^{y}}_t +\braket{\bar{\Pi}}_t \left[\kappa_{\mathrm{b}} -(\kappa_{\mathrm{b}}+\kappa_{\mathrm{c}})\braket{n}_t  \right] ~.
\label{SM:QCP-ct}
\end{align}
We note that the projector used in our model here is slightly  different from the one considered Ref.~\cite{Buchhold2017}, where $\bar{\Pi}_k = n_{k-1} + n_{k+1}$. This minor difference is not expected to modify the nonequilibrium behavior of the model. 

Discretising the above differential equation, and using that $\id = \Pi + \bar{\Pi}$ and $\id = n + \nb$, we find the equation 
\begin{align}
    \braket{n}_{t+1}&= (1 - \gamma \Delta t) \braket{\Pi}_{t} \braket{n}_{t} + \left(1-\gamma \Delta t-\kappa_{\mathrm{c}}\Delta t\right)\braket{\bar{\Pi}}_t\braket{n}_t+\kappa_{\mathrm{b}}\Delta t \braket{\bar{\Pi}}_t\braket{\bar{n}}_t
+ \Omega \Delta t\braket{\bar{\Pi}}_t \braket{\sigma^{y}}_{t} \, .
\label{SM:QCP-dt}
\end{align}

We note that by taking $\Omega\Delta t\to0$ in Eq.~\eqref{SM:QCP-ct} and in Eq.~\eqref{SM:QCP-dt}, one recovers the mean-field equations for the classical contact process in continuous-time and in a discrete-time approximation, respectively. In particular, we note that the discrete-time approximation of the dynamical equation of the classical contact process reads
$$
\braket{n}_{t+1}= (1 - \gamma \Delta t) \braket{\Pi}_{t} \braket{n}_{t} + \left(1-\gamma \Delta t-\kappa_{\mathrm{c}}\Delta t\right)\braket{\bar{\Pi}}_t\braket{n}_t+\kappa_{\mathrm{b}}\Delta t \braket{\bar{\Pi}}_t\braket{\bar{n}}_t\, . 
$$
Comparing this equation with Eq.~\eqref{eqn:full_qca_ccp_mfe} in the main text, we find how the rates of the continuous-time contact process can be obtained from the probabilities of our synchronous gate. We indeed find the coagulation rate $\kappa_{\mathrm{c}}=(q_{\circ\bullet\circ}-p_\bullet)/\Delta t$, the branching rate $\kappa_{\mathrm{b}}=p_{\circ}/\Delta t$ and decay rate $\gamma=p_{\circ\bullet\circ}/\Delta t$.
\end{document}